\documentclass[12pt]{article}
\usepackage{jhep-mod}
\usepackage{bm}
\usepackage{amssymb,amsmath,amsthm,physics}
\usepackage{mathrsfs}
\usepackage[utf8]{inputenc}
\usepackage{enumerate}
\hypersetup{colorlinks=true} 
\usepackage{xcolor}
\usepackage{appendix}
\usepackage{graphicx}
\usepackage{dcolumn}
\usepackage{bm}
\usepackage{multirow}
\usepackage{float}
\usepackage{tikz}
\usepackage{colortbl}
\usepackage[normalem]{ulem}
\definecolor{purple}{rgb}{1,0,1}
\definecolor{lime}{HTML}{A6CE39} 

\newcommand{\orcidicon}{%
	\begin{tikzpicture}
	\draw[lime, fill=lime] (0,0) 
		circle [radius=0.16] 
		node[white] {{\fontfamily{qag}\selectfont \tiny ID}};
	\draw[white, fill=white] (-0.0625,0.095) 
		circle [radius=0.007];
	\end{tikzpicture}	\hspace{-2mm}
}
\newcommand\orcidThomas{{\href{https://orcid.org/0000-0002-0314-4136}{\orcidicon}}}
\newcommand\orcidMatt{{\href{https://orcid.org/0000-0003-1088-6485}{\orcidicon}}}
\begin{document}

\title{\huge  
\leftline{Relativistic combination of non-collinear}
 3-velocities using quaternions\\
}

\author{
\Large Thomas Berry\orcidThomas  {\sf  and} Matt Visser\orcidMatt
}
\affiliation{
School of Mathematics and Statistics, Victoria University of Wellington, \\
\null\qquad PO Box 600, Wellington 6140, New Zealand}
\emailAdd{thomas.berry@sms.vuw.ac.nz;~matt.visser@sms.vuw.ac.nz}

\abstract{
\parindent0pt
\parskip7pt

\vspace{-10pt}
Quaternions have  an (over a century-old) extensive and quite complicated interaction with special relativity. Since quaternions are intrinsically 4-dimensional, and do such a good job of handling 3-dimensional rotations, the hope has always been that the use of quaternions would simplify some of the algebra of the Lorentz transformations. Herein we report a relatively nice result for the relativistic combination of non-collinear 3-velocities. If we work with the relativistic half-velocities $w$ defined by $v={2w\over1+w^2}$, and promote them to quaternions using $\vb{w} = w \; \vb{\hat n}$, where $\vb{\hat n}$ is a unit quaternion, then we shall show
\[
\vb{w}_{1\oplus2} = \vb{w}_1 \oplus \vb{w}_2
=(1-\vb{w}_1\vb{w}_2)^{-1} (\vb{w}_1 +\vb{w}_2)
= (\vb{w}_1 +\vb{w}_2)(1-\vb{w}_2\vb{w}_1)^{-1}.
\]
All of the complicated angular dependence for relativistic combination of non-collinear 3-velocities is now encoded in the quaternion multiplication of $\vb{w}_1$ with $\vb{w}_2$. 
This result can furthermore be extended to obtain an elegant and compact formula for the associated Wigner angle:
\[
\mathrm{e}^{\vb{\Omega}} = \mathrm{e}^{\Omega \; \vb{\hat\Omega} } 
= (1-\vb{w}_1\vb{w}_2)^{-1} (1-\vb{w}_2\vb{w}_1),
\]
in terms of which
\[
{\vb{\hat{n}}}_{1\oplus2} = \mathrm{e}^{\vb{\Omega}/2} \;\;
{\vb{w}_1+\vb{w}_2\over |\vb{w}_1+\vb{w}_2|}; 
\qquad\qquad
{\vb{\hat{n}}}_{2\oplus1} = \mathrm{e}^{-\vb{\Omega}/2} \;\;
{\vb{w}_1+\vb{w}_2\over |\vb{w}_1+\vb{w}_2|}.
\]
Thus, we would argue, many key results that are ultimately due to the non-commutativity of non-collinear boosts can be easily rephrased in terms of the algebra of quaternions.
 
\bigskip
{\sc Date:} 25 February 2020; 13 March 2020; \LaTeX-ed \today



{\sc PhySH:} \\
general physics; special relativity.
}

\maketitle


\def\tr{{\mathrm{tr}}}
\def\diag{{\mathrm{diag}}}
\newcommand{\eq}[1]{equation{#1}}			
\newcommand{\bb}[1]{\mathbb{#1}}			
\newcommand{\com}[2]{\left[#1, #2\right]}		
\newcommand{\w}[1]{\vb{w}_{#1}}			
\newcommand{\ii}{\vb{i}}					
\newcommand{\jj}{\vb{j}}					
\newcommand{\kk}{\vb{k}}					
\newcommand{\q}{\vb{q}}					
\renewcommand{\v}{\vec{v}}				
\newcommand{\e}{\mathrm{e}}		
\theoremstyle{definition}
\newtheorem{example}{Example}
\parindent0pt
\parskip7pt
\section{Introduction}
\label{S:introduction}

Hamilton first described the quaternions in the mid-1800s, primarily 
with a view to finding algebraically simple ways to handle 3-dimensional rotations.
With the advent of special relativity in 1905, and noting the manifestly 4-dimensional nature of quaternions once one adds a real part, multiple authors have tried to interpret special relativity in an intrinsically  quaternionic fashion~\cite{silberstein:1912, silberstein:1914, silberstein:wikipedia, dirac:1944, rastall:1964, girard:1984, ungar:1989, mocanu:1992}.

Despite technical success in applying quaternions to  special relativity, the use of quaternions in this subject has never really gained all that much traction in the physics community.
Perhaps one of the reasons for this is that there are a number of sub-optimal notational choices in Silberstein's original work~\cite{silberstein:1912, silberstein:1914, silberstein:wikipedia}, and the fact that there is no generally accepted way of using quaternions to represent Lorentz transformations, with many different authors employing their own quite distinct methods~\cite{silberstein:1912, silberstein:1914, silberstein:wikipedia, dirac:1944, rastall:1964, girard:1984, ungar:1989, mocanu:1992}. 

Below we shall introduce what we feel is a particularly simple and straightforward  method for combining relativistic 3-velocities using quaternions.
All of the interesting features due to non-commutativity properties of non-collinear boosts are implicitly and rather efficiently dealt with by the algebra of quaternions.
The method is based on an extension of an analysis by Giust, Vigoureux, and Lages~\cite{vigoureuxetal:2009,vigoureuxetal:2008}, who (because they were working with the usual complex numbers) were essentially limited to motion in 2-space;
their formalism is not really well-adapted to general motions in 3-space.

\section{Preliminaries}
\subsection{Lorentz transformations}
\enlargethispage{30pt}
The set of all Lorentz transformations of space-time form a group called the Lorentz group.
Mathematically, the Lorentz group is isomorphic to \( \text{O}(1,3) \), the orthogonal group of one time and three space dimensions that preserves the space-time interval
\begin{equation}
    -t^2+x^2+y^2+z^2.
\label{eq;stinterval}
\end{equation}
It is clear from this description that rotations of space-time are included in the Lorentz group, as well as the more familiar pure Lorentz transformations (boosts).
In fact, the pure Lorentz transformations do not even form a subgroup of the Lorentz group as, in general, the composition of two boosts \( B_1 \) and \( B_2 \) is not another boost but in fact a boost and a rotation \( B_{12}R_{12} = B_1 B_2\); while \( B_{21} R_{21} = B_2 B_1\).
This rotation, known as the Wigner rotation, was first discovered by Llewellyn Thomas in 1926 whilst trying to describe the Zeeman effect from a relativistic view-point~\cite{thomas:1926}, and was more fully analyzed by Eugene Wigner in 1939 \cite{wigner:1939}.
(For more recent discussions see~\cite{fisher:1972, ferraro:1999, malykin:2006, ritus:2007, visser_odonnell:2011}.)

It is well--known that the composition of Lorentz transformations is non-commutative.
That is, applying two successive boosts \( B_1 \) and \( B_2 \) in different orders results in the same final boost, \( B_{12} = B_{21} \), but different rotations, \( R_{12} \neq R_{21} \).
In the context of the combination of two velocities \( \v_1 \) and \( \v_2 \), this means that the final speed is the same no matter the order we combine the velocities, \( \norm{ \v_1 \oplus \v_2 } = \norm{ \v_2 \oplus \v_1 } \), but the final directions they point in are different $\hat v_{1\oplus2} \neq \hat v_{2\oplus 1}$.
Although not immediately obvious, the angle between \( \v_1 \oplus \v_2 \) and \( \v_2 \oplus \v_1 \) is in fact the Wigner angle \( \Omega \) \cite{visser_odonnell:2011}.
The Lorentz group has very many different representations, one of which is formulated by using the quaternions~\cite{silberstein:1912,dirac:1944}.

\subsection{Quaternions}		\label{sec;quaternions}

The quaternions are numbers that can be written in the form \( a+b\,\ii+c\,\jj+d\,\kk \), where \( a, \) \(b,\) \( c, \) and \( d \) are real numbers, and \( \ii, \) \( \jj, \) and \( \kk \) are the quaternion units which satisfy the famous relation 
\begin{equation}
    \ii^2 = \jj^2 = \kk^2 = \ii\jj\kk = -1.
\end{equation}
They form a four--dimensional number system that is generally treated as an extension of the complex numbers.
We shall define the quaternion conjugate of the quaternion \( \vb{q} = a + b\,\ii + c\,\jj + d\,\kk \) to be \( \vb{q}^\star = a - b\,\ii - c\,\jj - d\,\kk \), and define the norm of \( \vb{q} \) to be \( \vb{q q}^\star = \abs{\vb{q}}^2 = a^2+b^2+c^2+d^2 \in \bb{R} \).
This allows us to evaluate the quaternion inverse as $\q^{-1} = \q^\star/|\q|^2$. 

For current purposes we  focus our attention on pure quaternions.
That is, quaternions of the form \( a\,\ii + b\,\jj + c\,\kk \).
Many quaternion operations become much simpler when we are dealing with pure quaternions.
For example, the product of two pure quaternions \( \vb{p} \) and \( \vb{q} \) is given by \( \vb{p}\vb{q} = -\vec{p}\cdot\vec{q} + (\vec{p}\cross\vec{q})\cdot(\ii, \jj, \kk) \), where, in general, we shall set \( \vb{v} = \vec{v} \cdot (\ii,\jj,\kk) \).
From this, we obtain the useful relations 
\begin{equation}
    \com{\vb{p}}{\vb{q}} = 2 ( \vec{p}\cross\vec{q}) \cdot (\ii, \jj, \kk), \qq{and} \{ \vb{p},\vb{q} \} = -2\, \vec{p}\cdot\vec{q}.
\label{eq;pureqrelats}
\end{equation}
A notable consequence of \eqref{eq;pureqrelats} is \( \vb{q}^2 = -\vec{q}\cdot\vec{q} = -q^2 =  -\abs{\vb{q}}^2 \).
There is a natural isomorphism between the space of pure quaternions and \( \bb{R}^3 \) given by 
\begin{equation}
    \ii \mapsto \hat{x}, \quad \jj \mapsto \hat{y}, \quad \kk \mapsto \hat{z};
\label{eq;natiso}
\end{equation}
where \( \hat{x}, \, \hat{y}, \) and \( \hat{z} \) are the standard unit vectors in \( \bb{R}^3 \).

\enlargethispage{20pt}
One of the most common uses for quaternions today (2020) is in the computer graphics community, where they are used to compactly and efficiently generate rotations in 3-space.
Indeed, if \( \vb{q} = \cos(\theta/2) + \vu{u}\sin(\theta/2) \) is an arbitrary unit quaternion and \( \vb{v} \) is the image of a vector in \( \bb{R}^3 \) under the isomorphism \eqref{eq;natiso}, then the mapping \( \vb{v} \mapsto \vb{q}\vb{v}\vb{q}^{-1} \) rotates \( \vb{v} \) through an angle \( \theta \) about the axis defined by \( \vu{u} \).
The mapping \( \vb{v} \mapsto \vb{q}\vb{v}\vb{q}^{-1} \) is called \textit{quaternion conjugation} by \( \vb{q} \).

\section{Combining two 3-velocities}	\label{sec:3-velocities}
In the paper by Giust, Vigoureux, and Lages \cite{vigoureuxetal:2009}, see also~\cite{vigoureuxetal:2008}, a method is developed to compactly combine relativistic velocities in two space dimensions, and by extension, coplanar relativistic velocities in 3 space dimensions.
In the following subsection, we first provide a short summary of their approach, and then in the next subsection extend their method to general non-coplanar 3-velocities.

\subsection{Velocities in the ($x$,$y$)-plane}

The success of this Giust, Vigoureux, and Lages approach relies on the angle addition formula for the hyperbolic tangent function,
\begin{equation}
	\tanh(\xi_1 + \xi_2) = \frac{ \tanh{\xi_1} + \tanh{\xi_2} }{ 1 + \tanh{\xi_1}\tanh{\xi_2} }.
\label{eq;tanhadd}
\end{equation}
The tanh function is a natural choice for combining relativistic velocities since it is limited to the interval \( \left[-1 ,1\right] \).
Indeed, using the rapidity \( \xi \) defined by \( v=  \tanh(\xi)  \), we can easily combine collinear relativistic speeds using \eq{} \eqref{eq;tanhadd}.
In order to use this for the combination of non-collinear relativistic 2-velocities, we replace each 2-velocity \( \v \) by the complex number 
\begin{equation}
\label{E:w}
	V = \tanh({\xi}/{2}) \; \e^{i\varphi}.
\end{equation}
Here \( \xi \) is the rapidity of the velocity \( \v \), and \( \varphi \) gives the orientation of \( \v \) according to some observer in the plane defined by \( \v_1 \) and \( \v_2 \).
Giust, Vigoureux, and Lages then define the composition law \( \oplus \) for coplanar velocities \( \v_1 \) and \( \v_2 \) by 
\begin{equation}
        	W = \tanh\frac{\xi}{2} \;\e^{i\varphi_{1\oplus2}} = V_1 \oplus V_2 = \frac{V_1 + V_2}{1 + \overline{V_2} \, V_1} = \frac{ \tanh\frac{\xi_1}{2}\,\e^{i\varphi_1} + \tanh\frac{\xi_2}{2}\, \e^{i\varphi_2} }{ 1 + \tanh\frac{\xi_2}{2}\,\e^{-i\varphi_2} \tanh\frac{\xi_1}{2}\,\e^{i\varphi_1} },
\label{eq;oplus2d}
\end{equation}
where \( \overline{V} \) is the standard complex conjugate of \( V \).
By using \( \xi/2 \) instead of \( \xi \) in equations \eqref{E:w} and  \eqref{eq;oplus2d}, we are actually dealing with the ``relativistic half--velocities'', \( \tanh(\xi/2) \), 
where
\begin{equation}
w = \tanh(\xi/2) ; \qquad v = \tanh(\xi) = {2w\over 1+w^2}.
\end{equation}
Using \eq{s} \eqref{eq;tanhadd} and \eqref{eq;oplus2d} we can easily retrieve the real velocity from the half-velocity by using \( \oplus \) operator: \( v= \tanh\xi = \tanh{\xi/2} \oplus \tanh{\xi/2} = w\oplus w\).
In terms of the half velocities
\begin{equation}
 w_{1\oplus2} \;\e^{i\varphi_{1\oplus2}} 
 = \frac{ w_1\,\e^{i\varphi_1} + w_2\, \e^{i\varphi_2} }{ 1 + w_1 w_2\,\,\e^{i(\varphi_1-\varphi_2)} }.
\end{equation}

The \( \oplus \) addition law is non-commutative, which is most easily seen by first setting \( \theta = \varphi_2 - \varphi_1 \), then $\Omega = \varphi_{1\oplus2} - \varphi_{2\oplus1}$, and finally observing that the ratio 
\begin{equation}
    \e^{i\Omega} = \frac{ 1 + \tanh\frac{\xi_1}{2}\tanh\frac{\xi_2}{2}\e^{i\theta} }{ 1 + \tanh\frac{\xi_1}{2}\tanh\frac{\xi_2}{2}\e^{-i\theta} }
    = {1+w_1w_2 \e^{i\theta}\over 1+ w_1 w_2 \e^{-i\theta}}
\label{eq;noncomplanar}
\end{equation}
is not equal to unity for non--zero \( \theta \), meaning that \( \Omega=\varphi_{1\oplus2} - \varphi_{2\oplus1} \) is non-zero.

The angle \( \Omega=\varphi_{1\oplus2} - \varphi_{2\oplus1} \) is in fact the Wigner angle \( \Omega \), so an expression for this angle can be obtained by taking the real and imaginary parts of \eq{} \eqref{eq;noncomplanar}: 
\begin{equation}
    \tan\frac{\Omega}{2} = \frac{ \tanh\frac{ \xi_1 }{2} \,\tanh\frac{\xi_2}{2}\, \sin\theta}{1+\tanh\frac{\xi_1}{2}\, \tanh\frac{\xi_2}{2} \,\cos\theta} 
    = {w_1 w_2 \sin\theta\over1+w_1w_2\cos\theta}.
\label{eq;wigvig}
\end{equation}
This expression does not explicitly appear in 
reference~\cite{vigoureuxetal:2009} though something functionally equivalent, in the form $\Omega=2\; \hbox{arg}(1+w_1w_2e^{i\theta})$, appears in reference~\cite{vigoureuxetal:2008}.

The \( \oplus \) law can be applied to any number of coplanar velocities by iteration:
\begin{equation}
    W = (((V_1 \oplus V_{2} )\oplus \dots \oplus V_{n-1} )\oplus V_n).
\end{equation}
Thus it would be desirable to cleanly extend this formalism to general three-dimensional velocities.
Note that the order of composition is important, as we shall see in more detail below, the $\oplus$ operation is in general \emph{not} associative.

\subsection{General 3-velocities}

We now extend the result of Giust, Vigoureux, and Lages to arbitrary 3-velocities in three dimensions.

\subsubsection{Algorithm}

Suppose we have a velocity \( \v_i \) in the \((x,y)\)-plane, represented by the pure quaternion \( \vb{w}_i = \tanh(\xi_i/2) \vu{n}_i = \tanh(\xi_i/2)\;(\ii \cos\theta_i + \jj \sin\theta_i ) \). 
Using the rules for quaternion multiplication, we can write this as \( \vb{w}_i = \tanh(\xi_i/2)\;(\cos\theta_i + \kk \sin\theta_i) \ii \).
The term inside the brackets now looks very similar to what would be a natural extension of the exponential function to the quaternions, \( \e^{\kk\theta} = \cos\theta + \kk\sin\theta \).
To formalise this, we define the exponential of a quaternion \( \vb{q} \) by the power series 
\begin{equation}
    \e^{\vb{q}} = \sum_{k=0}^\infty \frac{\vb{q}^k}{k!}.
\label{eq;expqdefine}
\end{equation}
To calculate an explicit formula for \eq{} \eqref{eq;expqdefine}, we first consider the case of a pure quaternion \( \vb{u} \).
We know from section \ref{sec;quaternions} that for a pure quaternion we have \( \vb{u}^2 = -\abs{\vb{u}}^2 \), and so we find \( \vb{u}^3 = -\abs{\vb{u}}^2\vb{u}, \, \vb{u}^4 = \abs{\vb{u}}^4, \) and so on.
Thus, we can compute 
\begin{align}
    \e^{\vb{u}} \equiv \sum_{k=0}^\infty \frac{\vb{u}^k}{k!} 
    &= \left(1 - {1\over2!}\abs{\vb{u}}^2 + {1\over4!}\abs{\vb{u}}^4 - \dots\right) + {\vb{u}\over\abs{\vb{u}}}\left( \abs{\vb{u}} - {1\over3!}\abs{\vb{u}}^3 + {1\over5!}\abs{\vb{u}}^5 - \dots\right)	\notag \\
    &= \cos\abs{\vb{u}} + \vu{u}\sin\abs{\vb{u}}.	\label{eq;expquat}
\end{align}
\enlargethispage{15pt}
Following the same procedure above, we find the exponential of a pure unit quaternion \( \vu{u} \) and real number \( \phi \) to be
\begin{equation}
	\e^{\vu{u}\phi} = \cos\phi + \vu{u}\sin\phi.
\label{eq;expphiu}
\end{equation}
This nice result reflects the expression for the exponential of a complex number.

We can now extend this result to any arbitrary quaternion \( \vb{q} = a + \vb{u} \) by noting that the real number \( a \) commutes with all the terms in \( \vb{u} \), thereby allowing us to write \( \e^{\vb{q}} = \e^a \e^{\vb{u}} \), where \( \e^{\vb{u}} \) has the same form as \eq{} \eqref{eq;expquat}.
Explicitly,
\begin{equation}
    \e^{\vb{q}} = \e^a (\cos\abs{\vb{u}} + \vu{u} \; \sin\abs{\vb{u}} ).
\label{eq;expqgen}
\end{equation}

The exponential of a quaternion possesses many of the same properties as the exponential of a complex number.
Two particularly useful ones we use below are
\begin{equation}
    \left( \e^{\vu{u}\phi} \right)^\star = \e^{-\vu{u}\phi} = \cos\phi-\vu{u}\sin\phi, \qq{and} \abs{\e^{\vu{u}\phi}}=1.
\end{equation}

Using these results, we are now justified in writing
\begin{equation}
 \vb{w}_i = \tanh(\xi_i/2)\, \e^{\kk\theta_i} \, \ii = w_i \, \e^{\kk\theta_i} \, \ii 
\end{equation}
for our velocity in the \((x,y)\)-plane.

Building on this result, we now find it appropriate to define the \( \oplus \) operator for general 3-velocities, \( \vb{w}_1=w_1 \vb{\hat n_1} \) and \( \vb{w}_2= w_2 \vb{\hat n_2} \), by:
\begin{equation}
	\vb{w}_{1\oplus2} = \vb{w}_1 \oplus \vb{w}_2 = (1 - \vb{w}_1\vb{w}_2)^{-1}(\vb{w}_1 + \vb{w}_2).
\label{eq;oplusgen}
\end{equation}
The usefulness of this definition is best understood by looking at a few examples.

\subsubsection{Example: Parallel velocities}
We consider two parallel velocities \( \v_1 \) and \( \v_2 \) represented by the quaternions
\begin{equation}
    \w{1} = \tanh\frac{\xi_1}{2} \, \vb{\hat n}  \qq{and} \w{2} = \tanh\frac{\xi_2}{2} \, \vb{\hat n} ,
\end{equation}
respectively.
Our composition law \eqref{eq;oplusgen} then gives 
\begin{align}
	\w{1\oplus2} &= \left( 1 + \tanh\frac{\xi_1}{2}\,\tanh\frac{\xi_2}{2} \right)^{-1} \left( \tanh\frac{\xi_1}{2}\,  + \vb{\hat n} \tanh\frac{\xi_2}{2}\,  \vb{\hat n}  \right) \notag \\
	&= \frac{ \tanh\frac{\xi_1}{2} + \tanh\frac{\xi_2}{2} }{1 + \tanh\frac{\xi_1}{2} \,\tanh\frac{\xi_2}{2} } \; \vb{\hat n}  	\notag\\
	&= \tanh\left(\frac{\xi_1+\xi_2}{2}\right)  \; \vb{\hat n} , 	
	\label{eq;paraveloc}
\end{align}
which is equivalent to 
\begin{equation}
\w{1\oplus2} = {w_1+w_2\over1+w_1w_2} \; \vb{\hat n},
\end{equation}
and hence, also equivalent to 
the well--known result for the relativistic composition of two parallel velocities, \begin{equation}
    \v_1 \oplus \v_2 = \frac{v_1 + v_2}{1 + v_1 v_2} \; \hat{n}.
\label{eq;paralleladdv}
\end{equation}

\subsubsection{Example: Perpendicular velocities}
We now consider two perpendicular velocities given by 
\begin{equation}
    \w{1} = w_1 \ii , \quad \w{2} = w_2 \, \jj,
\end{equation}
where we have written \( \tanh(\xi_1/2) = w_1 \) and \( \tanh(\xi_2/2) = w_2 \) for brevity.

Our composition law then gives a combined velocity of 
\begin{equation}
    \vb{w}_{1\oplus2} = (1 - w_1w_2 \ii\jj)^{-1} (w_1\ii + w_2\jj) = \frac{w_1(1-w_2^2) \ii + w_2(1+w_1^2) \jj}{1+w_1^2w_2^2},
\end{equation}
which is definitely not commutative. In contrast the norm is symmetric:
\begin{equation}
    \abs{\vb{w}_{1\oplus2}}^2 = \frac{w_1^2(1-w_2^2)^2 + w_2^2(1+w_1^2)^2}{(1+w_1^2w_2^2)^2} = \frac{w_1^2+w_2^2+w_1^2w_2^4+w_2^2w_1^4}{(1+w_1^2w_2^2)^2} = \frac{w_1^2+w_2^2}{1+w_1^2w_2^2}.
\label{eq;w12perp}
\end{equation}
Here the \( \vb{w}_i \) are the ``relativistic half--velocities'' \( w_i = \tanh(\xi_i/2) \), so the full velocities are
\begin{equation}
    \abs{\vb{v}_i}^2 =  \abs{\vb{w}_i \oplus \vb{w}_i}^2 = \frac{4w_i^2}{(1+w_i^2)^2},
\label{eq;visquared}
\end{equation}
and so give a final speed of
\begin{equation}
    \abs{\vb{v}_{1\oplus2}}^2 = \frac{4\left(w_1^2+w_2^2\right)}{\big(1+w_1^2w_2^2\big) \left[1+\frac{w_1^2+w_2^2}{1+w_1^2w_2^2}\right]^2} = \frac{4\big(w_1^2+w_2^2\big) \big(1+w_1^2w_2^2\big)}{\left[ \big(1+w_1^2\big) \big(1+w_2^2\big) \right]^2}.
\end{equation}
The non-quaternionic result for the composition of two perpendicular velocities is \cite{visser_odonnell:2011}
\begin{equation}
    \norm{\v_{1\oplus2}}^2 = v_1^2 + v_2^2 - v_1^2v_2^2.
\end{equation}
Thus, we find
\begin{equation}
    \norm{\v_{1\oplus2}}^2 = \frac{4w_1^2}{(1+w_1^2)^2} + \frac{4w_2^2}{(1+w_2^2)^2} - \frac{16w_1^2w_2^2}{(1+w_1^2)^2(1+w_2^2)^2} = \frac{4(w_1^2+w_2^2)(1+w_1^2w_2^2)}{\left[ (1+w_1^2)(1+w_2^2) \right]^2}.
\end{equation}
And so our composition law \( \oplus \) gives the standard result for the composition of two perpendicular velocities.

\subsubsection{Example: Reduction to Giust--Vigoureux--Lages result}
It is important to note that our composition law \( \oplus \) reduces to the composition law of Giust, Vigoureux, and Lages when dealing with planar velocities.
As above, we define general velocities in the (\( \ii, \jj \))-plane by \( \w{1} = \tanh(\xi_1/2) \e^{\kk\phi_1} \ii \), and \( \w{2} = \tanh(\xi_2/2) \e^{\kk\phi_2} \ii \), then, using our composition law \eqref{eq;oplusgen}, we find
\begin{equation}
    \w{1\oplus2} = \left(1 - \tanh\frac{\xi_1}{2} \e^{\kk\phi_1}\ii \; \tanh\frac{\xi_2}{2} \e^{\kk\phi_2}\ii \right)^{-1} \left( \tanh\frac{\xi_1}{2} \e^{\kk\phi_1}\ii + \tanh\frac{\xi_2}{2} \e^{\kk\phi_2}\ii \right).
\end{equation}
But, noting that \( \tanh(\xi_2/2) \e^{\kk\phi_2}\ii = \tanh(\xi_2/2) \ii \, \e^{-\kk\phi_2}  \) and $\ii^2=-1$, we can re-write this as 
\begin{equation}
    \w{1\oplus2} = \left(1 + \tanh\frac{\xi_1}{2} \e^{\kk\phi_1} \tanh\frac{\xi_2}{2} \e^{-\kk\phi_2}\ \right)^{-1} \left( \tanh\frac{\xi_1}{2} \e^{\kk\phi_1} + \tanh\frac{\xi_2}{2} \e^{\kk\phi_2}\right)\ii.
\end{equation}
\enlargethispage{20pt}
Now writing 
\begin{equation}
\w{1\oplus2} = \tanh(\xi_{1\oplus2}/2) \, \e^{\kk \phi_{1\oplus2}}\,  \ii
\end{equation}
we can cancel out the trailing $\ii$, to obtain
\begin{equation}
 \tanh{\xi_{1\oplus2}\over2}\; \e^{\kk \phi_{1\oplus2}} =
 \left(1 + \tanh\frac{\xi_1}{2} \e^{\kk\phi_1} \tanh\frac{\xi_2}{2} \e^{-\kk\phi_2}\ \right)^{-1} \left( \tanh\frac{\xi_1}{2} \e^{\kk\phi_1} + \tanh\frac{\xi_2}{2} \e^{\kk\phi_2}\right).
\end{equation}
This expression now only contains $\kk$,  so everything commutes, and we can write
\begin{equation}
w_{1\oplus2}\,\e^{\kk \phi_{1\oplus2}} = 
 {
 w_1\, \e^{\kk\phi_1} + w_2\, \e^{\kk\phi_2}
 \over
 1 + w_1\, \e^{\kk\phi_1}\, w_2\, \e^{-\kk\phi_2}
 }
\end{equation}
which is equivalent to the result of Giust, Vigoureux, and Lages. 

\subsubsection{Uniqueness}
Finally, we might note that the expression for the composition law \eqref{eq;oplusgen} is not unique.
For example, by considering the power-series of \( (1-\w{1}\w{2})^{-1} \), we can re-write \eq{} \eqref{eq;oplusgen} as
\begin{equation}
    \w{1\oplus2} = (1-\w{1}\w{2})^{-1} (\w{1}+\w{2}) = \sum_{n=0}^\infty (\w{1}\w{2})^n (\w{1}+\w{2}).
\end{equation}
But, as \( \w{1} \) and \( \w{2} \) are pure quaternions, both \( \w{1}^2 \) and \( \w{2}^2 \) are real numbers, and so commute with \( \w{1} \) and \( \w{2} \).
Thus,
\begin{equation}
    \w{1\oplus2} = \sum_{n=0}^\infty (\w{1}\w{2})^n \w{1} + \sum_{n=0}^\infty (\w{1}\w{2})^n \w{2} = \w{1} \sum_{n=0}^\infty (\w{2}\w{1})^n + \w{2} \sum_{n=0}^\infty (\w{2}\w{1})^n.
\end{equation}
Consequently we find that our composition law can also be written as 
\begin{equation}
    \w{1\oplus2} = (\w{1}+\w{2})\sum_{n=0}^\infty (\w{2}\w{1})^n = (\w{1}+\w{2}) (1 - \w{2}\w{1})^{-1}.
\label{eq;oplusgen2}
\end{equation}
Indeed, one could use either \eq{} \eqref{eq;oplusgen} or \eq{} \eqref{eq;oplusgen2} as the definition of the composition law \( \oplus \).
Nonetheless, we will stick with the convention given in \eqref{eq;oplusgen}.

\subsection{Calculating the Wigner angle}

In this section we obtain an expression for the Wigner angle for general 3-velocities using our composition law \eqref{eq;oplusgen}.
Our calculations are obtained using the result that the Wigner angle is the angle between the velocities \( \w{1\oplus2} \) and \( \w{2\oplus1} \).

We first note
\begin{equation}
|\w{1\oplus2}| = |\w{2\oplus1}| = |1-\w{1}\w{2}|^{-1} |\w{1}+\w{2}| 
= {|w_1 \vb{\hat n}_1+w_2 \vb{\hat n}_2| \over |1- w_1 w_2 \; \vb{\hat n}_1 \vb{\hat n}_2|}. 
\end{equation}
Thence, setting $\cos\theta=\vec n_1 \cdot \vec n_2$ we explicitly verify
\begin{equation}
|\w{1\oplus2}|  = |\w{2\oplus1}|= \sqrt{w_1^2+w_2^2+2w_1 w_2 \cos\theta
 \over 1 + w_1^2 w_2^2 + 2 w_1w_2 \cos\theta}. 
\end{equation}
Now note that because $|\w{1\oplus2}| = |\w{2\oplus1}|$ it follows that  $\left(\w{1\oplus2}\right) \left(\w{2\oplus1}\right)^{-1}$ is a unit norm quaternion. 
In fact it is related to the Wigner angle by
\begin{equation}
\e^{\vb{\Omega}} =\left(\w{1\oplus2}\right) \left(\w{2\oplus1}\right)^{-1}.
\end{equation}
Then
\begin{equation}
\e^{\vb{\Omega}} = \left( (1-\w{1}\w{2})^{-1} (\w{1}+\w{2}) \right)
\left(  (1-\w{2}\w{1})^{-1} (\w{2}+\w{1}) \right)^{-1}
\end{equation}
But since for a product of quaternions $(\q_1\q_2)^{-1} = \q_2^{-1} \q_1^{-1}$ this reduces to
\begin{equation}
\e^{\vb{\Omega}} =  (1-\w{1}\w{2})^{-1}  (1-\w{2}\w{1}).
\end{equation}
Now
\begin{equation}
\w{1}\w{2} = - w_1w_2\cos\theta + (\vec w_1\times \vec w_2)\cdot(\ii,\jj,\kk).
\end{equation}
Let us define
\begin{equation}
{\hat\Omega} = {(\vec w_1\times \vec w_2)\over |\vec w_1\times \vec w_2|};
\qquad\hbox{so}\qquad
{\hat w_1}\times {\hat w_2}= \sin\theta \;\; {\hat\Omega}.
\end{equation}
Then
\begin{equation}
\w{1}\w{2} = - w_1w_2 (\cos\theta - \sin\theta\; \vb{\hat\Omega}) 
 =-w_1w_2 \;\e^{-\theta \vb{\hat\Omega}}.
\end{equation}
Consequently the Wigner angle satisfies
\begin{equation}
\e^{\vb{\Omega}} = \e^{\Omega \,\vb{\hat\Omega}} =  
\left(1+w_1w_2\, \e^{-\theta \vb{\hat\Omega}}\right)^{-1}  
\left(1+w_1w_2 \,\e^{\theta \vb{\hat\Omega}}\right)
= {1+w_1w_2 \,\e^{\theta \vb{\hat\Omega}}\over  1+w_1w_2 \,\e^{-\theta \vb{\hat\Omega}}}.
\end{equation} 
\enlargethispage{10pt}
Equivalently, 
\begin{equation}
	\e^{\Omega \vb{\hat\Omega}/2} 
	= \frac{1 + w_1w_2\,\e^{\theta\vb{\hat\Omega}}}{\abs{1 + w_1w_2 \, \e^{\theta\vb{\hat\Omega}}}}.
\label{eq;wig4}
\end{equation}
Taking the scalar and vectorial parts of \eq{} \eqref{eq;wig4}, we finally obtain 
\begin{equation}
	\tan\frac{\Omega}{2} = \frac{w_1w_2\sin\theta}{1+w_1w_2\cos\theta} = \frac{\abs{\vec{w}_1\cross \vec{w}_2}}{1+\vec{w}_1 \cdot \vec{w}_2},
\label{eq;tanwighalf}
\end{equation}
as an explicit expression for the Wigner angle \( \Omega \).

The simplicity of \eq{} \eqref{eq;tanwighalf} compared to exisiting formulae for \( \Omega \) in the literature, shows how the composition law \eqref{eq;oplusgen} can lead to much tidier and simpler formulae than other methods allowed for.
This can be seen as the extension of the result \eqref{eq;wigvig} to more general velocities.

We can write \eq{} \eqref{eq;tanwighalf} in a perhaps more familiar (though possibly more tedious) form by first noting that from \eq{} \eqref{eq;visquared} we have
\begin{equation}
    w_i = \frac{1-\sqrt{1-v_i^2}}{v_i} = \frac{\gamma_i-1}{\sqrt{\gamma_i^2-1}} = \sqrt{ \frac{\gamma_i-1}{\gamma_i+1} } = \frac{ \sqrt{\gamma_i^2-1} }{\gamma_i + 1} = \frac{ v_i \gamma_i }{\gamma_i+1},
\end{equation}
and so
\begin{equation}
    \tan\frac{\Omega}{2} = \frac{v_1v_2 \gamma_1\gamma_2 \sin\theta}{(1+\gamma_1)(1+\gamma_2) + v_1v_2\gamma_1\gamma_2\cos\theta}.
\label{eq;tanhalfreal}
\end{equation}

We can check two interesting cases of \eq{} \eqref{eq;tanwighalf} for when \( \theta = 0 \) (parallel velocities) and when \( \theta = \pi/2 \) (perpendicular velocities).
We can see directly that, for parallel velocities, the associated Wigner angle is given by \( \tan(\Omega/2) = 0 \), so that \( \Omega = n\pi \) for \( n \in \bb{Z} \); whilst for perpendicular velocities, the associated Wigner angle is simply given by \( \tan(\Omega/2) = w_1w_2 \).

It is easiest to check our results against the literature using the somewhat messier \eq{} \eqref{eq;tanhalfreal}, in which case parallel velocities again give \( \tan(\Omega/2) = 0 \), whilst perpendicular velocities give 
\begin{equation}
    \tan(\Omega/2) = \frac{v_1v_2\gamma_1\gamma_2}{(1+\gamma_1)(1+\gamma_2)},
\end{equation}
which agrees with the results given in \cite{visser_odonnell:2011}.

\section{Combining three 3-velocities}	\label{sec:three-3-velocities}
Let us now see what happens when we relativistically combine 3 half-velocities.\\
We shall calculate, compare, and contrast $\vb{w}_{(1\oplus2)\oplus3 }$ 
with $\vb{w}_{1\oplus(2\oplus3)}$.

\subsection{Combining 3 half-velocities: $\vb{w}_{(1\oplus2)\oplus3 }$}
Start from our key result
\begin{equation}
\vb{w}_{1\oplus2} = \vb{w}_1 \oplus \vb{w}_2
=(1-\vb{w}_1\vb{w}_2)^{-1} (\vb{w}_1 +\vb{w}_2),
\end{equation}
and iterate it to yield
\begin{equation}
\vb{w}_{(1\oplus2)\oplus3 } 
= \{1-(1-\vb{w}_1\vb{w}_2)^{-1} (\vb{w}_1 +\vb{w}_2)\vb{w}_3 \}^{-1} 
\{(1-\vb{w}_1\vb{w}_2)^{-1} (\vb{w}_1 +\vb{w}_2)+ \vb{w}_3\}.
\end{equation}

It is now a matter of straightforward quaternionic algebra to check that
\begin{eqnarray}
\vb{w}_{(1\oplus2)\oplus3 } 
&=&
\{(1-\vb{w}_1\vb{w}_2)^{-1} (1-\vb{w}_1\vb{w}_2 -(\vb{w}_1 +\vb{w}_2)\vb{w}_3 )\}^{-1} 
\nonumber\\
&& \qquad\qquad \times
\{(1-\vb{w}_1\vb{w}_2)^{-1} (\vb{w}_1 +\vb{w}_2)+ \vb{w}_3\}
\nonumber\\
&=&
(1-\vb{w}_1\vb{w}_2 -(\vb{w}_1 +\vb{w}_2)\vb{w}_3 )^{-1} (1-\vb{w}_1\vb{w}_2) \{(1-\vb{w}_1\vb{w}_2)^{-1} (\vb{w}_1 +\vb{w}_2)+ \vb{w}_3\}
\nonumber\\
&=&
(1-\vb{w}_1\vb{w}_2 -(\vb{w}_1 +\vb{w}_2)\vb{w}_3 )^{-1}  \{(\vb{w}_1 +\vb{w}_2)+ (1-\vb{w}_1\vb{w}_2) \vb{w}_3\}.
\end{eqnarray}
Ultimately
\begin{equation}
\vb{w}_{(1\oplus2)\oplus3 } 
=  \{1-\vb{w}_1\vb{w}_2 -\vb{w}_1\vb{w}_3- \vb{w}_2\vb{w}_3 \}^{-1}  \{\vb{w}_1 +\vb{w}_2+ \vb{w}_3 - \vb{w}_1\vb{w}_2 \vb{w}_3\}.
\label{E:(12)3a}
\end{equation}

An alternative formulation starts from
\begin{equation}
\vb{w}_{1\oplus2} = \vb{w}_1 \oplus \vb{w}_2
= (\vb{w}_1 +\vb{w}_2) (1-\vb{w}_2\vb{w}_1)^{-1},
\end{equation}
which when iterated yields
\begin{equation}
\vb{w}_{(1\oplus2)\oplus3 } 
= 
\{(\vb{w}_1 +\vb{w}_2) (1-\vb{w}_2\vb{w}_1)^{-1} + \vb{w}_3\}
\{1-  \vb{w}_3 (\vb{w}_1 +\vb{w}_2) (1-\vb{w}_2\vb{w}_1)^{-1} \}^{-1}.
\end{equation}
Thence a little straightforward quaternionic algebra verifies that
\begin{eqnarray}
\vb{w}_{(1\oplus2)\oplus3 } 
&=& 
\{(\vb{w}_1 +\vb{w}_2)  + \vb{w}_3(1-\vb{w}_2\vb{w}_1)\}(1-\vb{w}_2\vb{w}_1)^{-1}
\nonumber\\
&&\qquad\qquad\times \{1-  \vb{w}_3 (\vb{w}_1 +\vb{w}_2) (1-\vb{w}_2\vb{w}_1)^{-1} \}^{-1}
\nonumber\\
&=&
\{(\vb{w}_1 +\vb{w}_2)  + \vb{w}_3(1-\vb{w}_2\vb{w}_1)\}
\{(1-\vb{w}_2\vb{w}_1)-  \vb{w}_3 (\vb{w}_1 +\vb{w}_2) \}^{-1}
\end{eqnarray}
Ultimately
\begin{equation}
\vb{w}_{(1\oplus2)\oplus3 } 
= 
\{\vb{w}_1 +\vb{w}_2  + \vb{w}_3 -\vb{w}_3\vb{w}_2\vb{w}_1\}
\{1-\vb{w}_2\vb{w}_1-  \vb{w}_3 \vb{w}_1 - \vb{w}_3 \vb{w}_2 \}^{-1}.
\label{E:(12)3b}
\end{equation}
So we have found two equivalent formulae for $\vb{w}_{(1\oplus2)\oplus3}$,
equations (\ref{E:(12)3a}) and (\ref{E:(12)3b}).

\subsection{Combining 3 half-velocities: $\vb{w}_{1\oplus(2\oplus3) }$}
In contrast, the situation for $\vb{w}_{1\oplus(2\oplus3) }$ is considerably more subtle.
Start from the key result that 
\begin{equation}
\vb{w}_{2\oplus3} = \vb{w}_2 \oplus \vb{w}_3
= (1-\vb{w}_2\vb{w}_3)^{-1} (\vb{w}_2 +\vb{w}_3),
\end{equation}
and iterate it to yield
\begin{equation}
\vb{w}_{1\oplus(2\oplus3) } 
=
\{1- \vb{w}_1(1-\vb{w}_2\vb{w}_3)^{-1} (\vb{w}_2 +\vb{w}_3)\}^{-1}
\{ \vb{w}_1 + (1-\vb{w}_2\vb{w}_3)^{-1} (\vb{w}_2 +\vb{w}_3)\}.
\end{equation}
The relevant quaternionic algebra is now a little trickier
\begin{eqnarray}
\vb{w}_{1\oplus(2\oplus3) } 
&=&
\{1- \vb{w}_1(1-\vb{w}_2\vb{w}_3)^{-1} (\vb{w}_2 +\vb{w}_3)\}^{-1}
(1-\vb{w}_2\vb{w}_3)^{-1}
\nonumber\\
&&\qquad\qquad \times\{ (1-\vb{w}_2\vb{w}_3)\vb{w}_1 +  (\vb{w}_2 +\vb{w}_3)\}
\nonumber\\
&=&
\{ (1-\vb{w}_2\vb{w}_3)(1- \vb{w}_1(1-\vb{w}_2\vb{w}_3)^{-1} (\vb{w}_2 +\vb{w}_3)\}^{-1}
\nonumber\\
&&\qquad\qquad\times
\{ (1-\vb{w}_2\vb{w}_3)\vb{w}_1 +  (\vb{w}_2 +\vb{w}_3)\}
\nonumber\\
&=&
\{ 1-\vb{w}_2\vb{w}_3- (1-\vb{w}_2\vb{w}_3) \vb{w}_1(1-\vb{w}_2\vb{w}_3)^{-1} (\vb{w}_2 +\vb{w}_3)\}^{-1}
\nonumber\\
&&\qquad\qquad\times
\{ \vb{w}_1 +  \vb{w}_2 +\vb{w}_3 - \vb{w}_2\vb{w}_3\vb{w}_1\}.
\end{eqnarray}
To proceed we note that
\begin{eqnarray}
(1-\vb{w}_2\vb{w}_3) \vb{w}_1(1-\vb{w}_2\vb{w}_3)^{-1}
&=&
\left(1-\vb{w}_2\vb{w}_3\over |1-\vb{w}_2\vb{w}_3|\right) 
\vb{w}_1
\left(1-\vb{w}_2\vb{w}_3\over |1-\vb{w}_2\vb{w}_3|\right)^{-1}
\nonumber\\
&&
\vphantom{\Bigg|} = e^{-\vb{\Omega}_{2\oplus3}/2} \;\vb{w}_1\; e^{+\vb{\Omega}_{2\oplus3}/2}.
\end{eqnarray}
Thence
\begin{equation}
\vb{w}_{1\oplus(2\oplus3) } 
= 
\{ 1-\vb{w}_2\vb{w}_3-  (e^{-\vb{\Omega}_{2\oplus3}/2} \vb{w}_1e^{+\vb{\Omega}_{2\oplus3}/2}) (\vb{w}_2 +\vb{w}_3)\}^{-1}
\{ \vb{w}_1 +  \vb{w}_2 +\vb{w}_3 - \vb{w}_2\vb{w}_3\vb{w}_1\}.
\label{E:1(23)}
\end{equation}
While structurally similar to the formulae  (\ref{E:(12)3a}) and (\ref{E:(12)3b}) for $\vb{w}_{(1\oplus2)\oplus3}$ the present result (\ref{E:1(23)}) for $\vb{w}_{1\oplus(2\oplus3)}$ is certainly different --- the Wigner angle $\vb{\Omega}_{2\oplus3}$ now makes an explicit appearance, also the form of the triple-product $\vb{w}_2\vb{w}_3\vb{w}_1$ is different.

\subsection{Combining 3 half-velocities: (Non)-associativity}
From  (\ref{E:(12)3a}) and (\ref{E:(12)3b}) for $\vb{w}_{(1\oplus2)\oplus3}$, and (\ref{E:1(23)}) for $\vb{w}_{1\oplus(2\oplus3)}$, it is clear that relativistic composition of velocities is in general not associative. (See for instance the discussion in references~\cite{ungar:2006, sonego:2006}, commenting on reference~\cite{sonego:2005}.) 

A sufficient condition for associativity,
 $\vb{w}_{(1\oplus2)\oplus3 } = \vb{w}_{1\oplus(2\oplus3) }$,   is to enforce
\begin{equation}
e^{-\vb{\Omega}_{2\oplus3}/2} \vb{w}_1e^{+\vb{\Omega}_{2\oplus3}/2} =  \vb{w}_1;
 \qquad\hbox{and} \qquad \vb{w}_1\vb{w}_2 \vb{w}_3 = \vb{w}_2\vb{w}_3\vb{w}_1.
\end{equation}
 That is, a sufficient condition for associativity is
 \begin{equation}
 [\vb{\Omega}_{2\oplus3}, \vb{w}_1]=0; \qquad\hbox{and} \qquad 
 [\vb{w}_1,\vb{w}_2 \vb{w}_3] = 0.
\end{equation}
But note $\vb{\Omega}_{2\oplus3}\propto [\vb{w}_2, \vb{w}_3]$ and $\vb{w}_2 \vb{w}_3 =
\{\vb{w}_2, \vb{w}_3\} +[\vb{w}_2, \vb{w}_3]$.
This now implies that these two sufficiency conditions are in fact identical; a sufficient condition for associativity is
\begin{equation}
 [\vb{w}_1,[\vb{w}_2, \vb{w}_3]] = 0.
\end{equation}
 This sufficient condition for associativity can also be written as the vanishing of the vector triple product
\begin{equation}
\vec w_1 \times(\vec w_2 \times \vec w_3) = 0.
\end{equation}

\subsection{Specific non-coplanar example}
As a final example of the power of the quaternion formalism, let us consider a specific  intrinsically  non-coplanar example. Let \( \w{1} = w_1 \ii \), \( \w{2} = w_2\, \jj \), and \( \w{3} = w_3 \kk \) be three mutually perpendicular half-velocities. (So this configuration does automatically satisfy the associativity condition discussed above.)

Then we have already seen that:
\begin{equation}
\w{1} \oplus \w{2} = \frac{w_1(1-w_2^2)\ii + w_2(1+w_1^2)\jj}{1+w_1^2w_2^2};
\qquad
w_{1\oplus2}^2 = {w_1^2 + w_2^2\over 1+w_1^2 w_2^2}.
\end{equation}
Furthermore, since $\w{1} \oplus \w{2}$ is perpendicular to $\w{3}$, we have
\begin{equation}
(\w{1} \oplus \w{2})\oplus \w{3} 
 = \frac{w_{1\oplus2}(1-w_3^2)\vb{\hat n}_{1\oplus2} + w_3(1+w_{1\oplus2}^2)\kk}{1+w_{1\oplus2}^2w_3^2},
\end{equation}
and
\begin{equation}
w_{(1\oplus2)\oplus3}^2 = {w_{(1\oplus2)}^2+w_3^2\over 1 + w_{(1\oplus2)}^2 w_3^2}
=
{w_1^2+w_2^2+w_3^2+w_1^2w_2^2w_3^2
\over 1 + w_1^2 w_2^2 + w_2^2 w_3^2 + w_3^2 w_1^2}.
\end{equation}

A little algebra now yields the manifestly non-commutative result
\begin{equation}
(\w{1}\oplus\w{2})\oplus\w{3} =
{
 (1-w_2^2)(1-w_3^2)\w{1}  + (1+w_1^2)(1-w_3^2)\w{2} + (1+w_1^2)(1+w_2^2)\w{3} 
\over
1+w_1^2w_2^2 +  w_2^2w_3^2 +w_3^2w_1^2
}.
\end{equation}
In this particular case we can also explicitly show that
\begin{equation}
(\w{1}\oplus\w{2})\oplus\w{3} = \w{1}\oplus(\w{2}\oplus\w{3}),
\end{equation}
though (as discussed above) associativity fails in general.

\section{Conclusions}

Herein we have provided a simple and elegant algebraic method for combining special relativistic 3-velocities using quaternions:
\begin{equation}
\vb{w}_{1\oplus2} = \vb{w}_1 \oplus \vb{w}_2
=(1-\vb{w}_1\vb{w}_2)^{-1} (\vb{w}_1 +\vb{w}_2)
= (\vb{w}_1 +\vb{w}_2)(1-\vb{w}_2\vb{w}_1)^{-1}.
\end{equation}
The construction also leads to an elegant formula for the Wigner angle:
\begin{equation}
e^{\vb{\Omega}} = e^{\Omega \; \vb{\hat\Omega} } = (1-\vb{w}_1\vb{w}_2)^{-1} (1-\vb{w}_2\vb{w}_1),
\end{equation}
in terms of which
\[
{\vb{\hat{n}}}_{1\oplus2} = \mathrm{e}^{\vb{\Omega}/2} \;\;
{\vb{w}_1+\vb{w}_2\over |\vb{w}_1+\vb{w}_2|}; 
\qquad\qquad
{\vb{\hat{n}}}_{2\oplus1} = \mathrm{e}^{-\vb{\Omega}/2} \;\;
{\vb{w}_1+\vb{w}_2\over |\vb{w}_1+\vb{w}_2|}.
\]
All of the non-commutativity associated with non-collinearity of 3-velocities is automatically and rather efficiently dealt with by the quaternion algebra.

\acknowledgments{
TB was supported by a Victoria University of Wellington MSc scholarship, and was also indirectly supported by the Marsden Fund, via a grant administered by the Royal Society of New Zealand.
MV was directly supported by the Marsden Fund, via a grant administered by the Royal Society of New Zealand.
The authors wish to thank Jos\'e Lages for useful comments.}



\begin{thebibliography}{69}  

\bibitem{silberstein:1912}
L~Silberstein,
\newblock ``Quaternionic form of relativity'',\\
\newblock {Philisophical Magazine}, {\bf 23 \# 137} (May 1912) 790--809.

\bibitem{silberstein:1914}
L~Silberstein,
\newblock \emph{The theory of relativity},
\newblock (Macmillan and Co, London, 1914)

\bibitem{silberstein:wikipedia}
See for example: \url{https://en.wikipedia.org/wiki/Ludwik_Silberstein}

\bibitem{dirac:1944}
P~A~M Dirac,
\newblock ``Application of quaternions to {Lorentz} transformations'',\\
\newblock {{Proceedings of the Royal Irish Academy. \\
Section A: Mathematical
  and Physical Sciences}}, {\bf 50} (1944) 261--270.
  
\bibitem{rastall:1964}
  P~Rastall,
  ``Quaternions in Relativity'',
  Rev.\ Mod.\ Phys.\  {\bf 36 \# 3} (1964)  820.
  doi:10.1103/RevModPhys.36.820
  
\bibitem{girard:1984}
P~R Girard,
\newblock ``The quaternion group and modern physics'',\\
\newblock {European Journal of Physics} {\bf 5 \# 1} (1984) 25--32.

\bibitem{ungar:1989}
A~A Ungar,
\newblock {``The relativistic velocity composition paradox and the Thomas
  rotation''},
\newblock {Foundations of Physics} {\bf 19 \# 11} (November 1989)  1385--1396.

\bibitem{mocanu:1992}
C~I Mocanu,\\
\newblock {``On the relativistic velocity composition paradox and the Thomas
  rotation''},\\
\newblock {Foundations of Physics Letters} {\bf 5 \# 5} (October 1992) 443--456.

\bibitem{vigoureuxetal:2009}
R~Giust, J-M~Vigoureux, and J~Lages,\\
\newblock ``Generalized composition law from $2 \times 2$ matrices'',\\
\newblock {American Journal of Physics} {\bf77 \# 11} (2009) 1068--1073.

\bibitem{vigoureuxetal:2008}
J~Lages, R~Giust, and J-M~Vigoureux,\\
``Composition law for polarizers'',
Physical Review A {\bf78} (2008) 033810\\
doi:	10.1103/PhysRevA.78.033810
[arXiv:0808.1355 [physics.optics]]

\bibitem{thomas:1926}
L~H Thomas,
\newblock ``The motion of the spinning electron'',\\
\newblock {Nature} {\bf 117 \# 2945} (1926) 514--514.

\bibitem{wigner:1939}
E~Wigner,
\newblock ``On unitary representations of the inhomogeneous {Lorentz} group'',\\
\newblock {Annals of Mathematics} {\bf40 \# 1} (1939) 149--204.


\bibitem{fisher:1972}
G~P~Fisher,
``Thomas precession'',
{American Journal of Physics} {\bf 40} (1972) 1772.

\bibitem{ferraro:1999}
M~Ferraro, R~Thibeault,\\
``Generic composition of boosts: an elementary derivation of the Wigner
  rotation'',\\
{European Journal of Physics} {\bf 20} (1999) 143.

\bibitem{malykin:2006}
G~B~Malykin,
``Thomas precession: correct and incorrect solutions'',\\
{Physics--Uspekhi} {\bf 49} (2006) 837--853.

\bibitem{ritus:2007}
V~I~Ritus,
``On the difference between Wigner's and M\o{}ller's approaches to the
  description of Thomas precession.''
{Physics--Uspekhi} {\bf 50} (2007) 95--101.

\bibitem{visser_odonnell:2011}
K~O'Donnell and M~Visser,
\newblock {``Elementary analysis of the special relativistic combination of
  velocities, Wigner rotation, and Thomas precession''},\\
\newblock { {European Journal of Physics}} {\bf 32} (2011) 1033--1047.

\bibitem{ungar:2006}
Abraham A Ungar,
``Thomas precession: a kinematic effect of the algebra of Einstein's velocity addition law. 
Comments on
\\
 `Deriving relativistic momentum and energy: II. Three-dimensional case'{}'',
\\
European Journal of Physics {\bf 27 \#3} (2006) L17.

\bibitem{sonego:2006}
Sebastiano Sonego and Massimo Pin,\\
``Deriving relativistic momentum and energy: II. Three-dimensional case (CORRIGENDUM)'',
European Journal of Physics {\bf27} (2006) 685.

\bibitem{sonego:2005}
Sebastiano Sonego and Massimo Pin,\\
``Deriving relativistic momentum and energy: II. Three-dimensional case'',\\
European Journal of Physics {\bf 25 \#5} (2005) 851--856.

\end{thebibliography}
\end{document}